\documentclass{emulateapj}
\usepackage{apjfonts}
\usepackage{graphicx}
\usepackage[dvips]{color}
\shorttitle{Blue Luminescence in the Red Rectangle}
\shortauthors{Vijh et al.}

\begin{document}

\title{Discovery of Blue Luminescence in the Red Rectangle: Possible Fluorescence from Neutral Polycyclic Aromatic Hydrocarbon Molecules?}

\author{Uma P. Vijh \altaffilmark{1,2}, Adolf N. Witt \altaffilmark{1,2} and Karl D. Gordon \altaffilmark{3}}
%\affil{}  
%\affil{Ritter Astrophysical Research Center}
%\email{uvijh@physics.utoledo.edu}
%\author{Adolf N. Witt \altaffilmark{1,2}}
%\affil{Ritter Astrophysical Research Center, University of Toledo, Toledo OH 43606}
%\email{awitt@dusty.astro.utoledo.edu}
%\and
%\author{Karl D. Gordon \altaffilmark{3}}
%\affil{}
%\affil{Steward Observatory}
%\email{kgordon@as.arizona.edu}

\altaffiltext{1}{Ritter Astrophysical Research Center, University of Toledo, Toledo, OH 43606}
\altaffiltext{2}{Visiting Astronomer, Cerro Tololo Inter-American Observatory.
CTIO is operated by AURA, Inc.\ under contract to the National Science
Foundation.}
\altaffiltext{3}{Steward Observatory, University of Arizona,Tucson, Arizona 85721}

\begin{abstract}
Here we report our discovery of a band of blue luminescence (BL) in the Red Rectangle (RR) nebula. This enigmatic proto-planetary nebula is also one of the brightest known sources of extended red emission as well as of unidentified infra-red (UIR) band emissions. The spectrum of this newly discovered BL is most likely fluorescence from small neutral polycyclic aromatic hydrocarbon (PAH) molecules. PAH molecules are thought to be widely present in many interstellar and circumstellar environments in our galaxy as well as in other galaxies, and are considered likely carriers of the UIR-band emission. However, no specific PAH molecule has yet been identified in a source outside the solar system, as the set of mid-infra-red emission features attributed to these molecules between the wavelengths of 3.3 $\mu$m and 16.4 $\mu$m is largely insensitive to molecular sizes. In contrast, near-UV/blue fluorescence of PAHs is more specific as to size, structure, and charge state of a PAH molecule. If the carriers of this near-UV/blue fluorescence are PAHs, they are most likely neutral PAH molecules consisting of 3-4 aromatic rings such as anthracene (C$_{14}$H$_{10}$) and pyrene (C$_{16}$H$_{10}$). These small PAHs would then be the largest molecules specifically identified in the interstellar medium.
\end{abstract}


\section{INTRODUCTION}
The family of emission bands at 3.3, 6.2, 7.7, 8.6, 11.2, \& 12.7 $\mu$m (called the unidentified infra-red (UIR) bands) is found in almost all astrophysical environments including the diffuse interstellar medium (ISM), the edges of molecular clouds, reflection nebulae, young stellar objects, HII regions, star forming regions, some C-rich Wolf-Rayet stars, post-AGB stars, planetary nebulae, novae, normal galaxies, starburst galaxies, most ultra-luminous infra-red galaxies and AGNs \citep[see][and references therein]{pet04}. Approximately 20-30\% of the Galactic IR radiation is emitted in these UIR bands and 10-15\% of the interstellar carbon is contained in the UIR carriers \citep{sw95}, indicating that the carriers represent an abundant component of the ISM. These UIR bands are the signatures of aromatic C-C and C-H fundamental vibrational and bending modes, and are generally attributed to a family of PAH molecules containing 50-100 carbon atoms \citep{hony01,cook98,lp84,ver01,atb85,sel84}. Although the presence of PAHs in space is widely accepted  by most astronomers, their specific sizes  and ionization states remain elusive. On absorption of a far-UV photon, a PAH molecule usually undergoes a transition to an upper electronic state. If the molecule undergoes iso-energetic transitions to highly vibrationally excited levels of the ground state, then the molecule relaxes through a series of IR emissions in the C-C and C-H vibrational and bending modes. These transitions are largely independent of size, structure and ionization state of the molecule.  However, \emph{electronic} fluorescence, a transition from the upper excited level to the ground state, is more specific \citep{reyle00}. In particular, the wavelength of the first electronic transition seen in neutral PAHs is closely dependent on the size of the molecular species. In general, this fluorescence wavelength increases with the molecular weight of the molecule (Figure 1). An observation of fluorescence in an astronomical source in the UV/visible range offers the possibility of estimating the size of the PAH molecules, which the observation of the UIR band emission in the same source does not.
%place fig1 here
\begin{figure}
\plotone{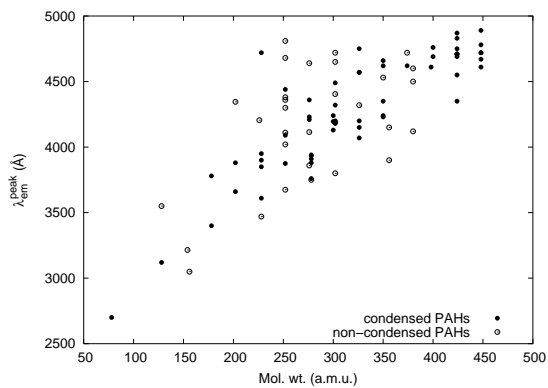}
\caption{The first electronic transition in the fluorescence spectra \citep{berl65,peaden80,reidel1,reidel2} of neutral PAH molecules as a function of their molecular weight. We distinguish between condensed and non-condensed PAHs, with condensed PAHs considered to be more stable under interstellar conditions.\label{fig1}}
\end{figure}

\section{TARGET AND OBSERVATIONS}
The Red Rectangle (RR) \citep{cohen75,van03} is a unique proto-planetary nebula. Its bipolar structure is a result of mass loss from an evolved central binary star, HD44179, directed by a circumbinary disk. Different aspects of the nebula become apparent in images at different wavelength regions shown in Figure 2. At blue wavelengths (3800 \AA\ $< \lambda <$ 4900 \AA\, bottom panel) the nebular structure is dominated by a bright spherical blob of about 8'' diameter, embedded in a faint rectangular envelope with approximate dimensions of 12'' x 24''. The blue spectrum of the nebula is dominated by dust-scattered light of the central A-type star. At red wavelengths (6220 \AA) the nebula appears to have sharp radial structures, representing the projected walls of a bi-conical outflow cavity, as shown in the high-resolution Hubble Space Telescope WFPC2 image in the middle panel. The nebular spectrum in the red region ($\lambda >$ 5400 \AA) is dominated by Extended Red Emission (ERE) \citep{wv04,wb90}, following a spatial distribution \citep{sw91} totally different from that of the dust-scattered light at blue wavelengths ($\lambda <$ 5000 \AA). The carrier of the ERE is still under investigation \citep{wv04}. The top panel shows a high-resolution short-exposure WFPC2 near-IR image of the central source, demonstrating in fact that the star itself is not directly visible from Earth. Instead, we see two blobs of scattered light above and below a disc \citep{buj03}, which obscures the star at visible wavelengths.

The RR nebula has the distinction of being the brightest known source of UIR band emission \citep{rsw78,geballe85} which is attributed to PAHs. In addition, the central star HD44179 is thought to be a post-AGB star in its first stage of evolution toward a planetary nebula. It is in this stage that substantial mass-loss takes place and a bi-polar structure develops. This stage in stellar evolution is considered a stage of active dust-production \citep{whit03}. Therefore, the RR nebula is a likely location in which to find observational evidence for small PAH molecules, which are the building blocks for larger PAH structures \citep{keller87} and carbon grains. In most other sources of UIR band emission, the emitting particles are a component of \emph{interstellar} dust, which is likely to have experienced substantial processing by UV radiation, interstellar shocks and chemical modification over a very long time span of existence since leaving the environment of its original formation. It is well established that PAH molecules with less than 40 C-atoms are unlikely to survive the harsh radiation conditions of interstellar space \citep{jbl99}, while they may well be abundant in the more benign, UV-poor radiation field of HD 44179, especially when partially shielded by the opaque disk apparent in Figure 2.
%place fig2 here
\begin{figure}
\begin{center}
\includegraphics[height=5in]{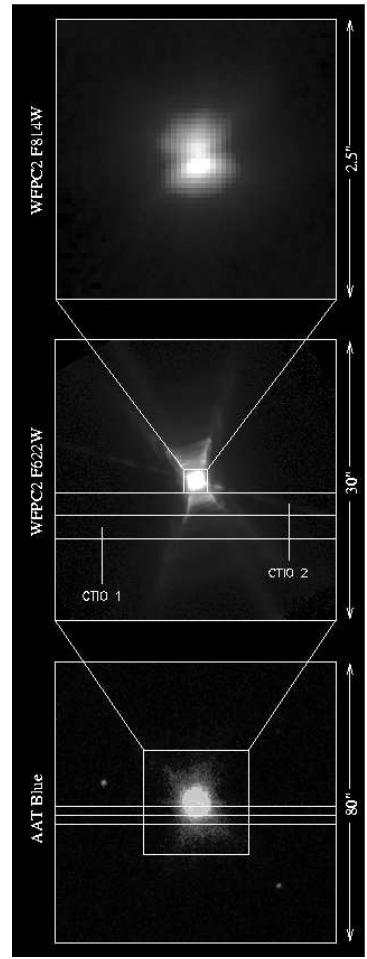}
\caption{The RR nebula. These images illustrate the observable structures of the RR nebula at different optical wavelengths and different angular scales. The bottom panel shows a blue image (3800 \AA\ - 4900 \AA), obtained using the 4-m AAT. The middle panel shows a HST WFPC2 image centered at a wavelength of 6220 \AA. The geometry of the central illuminating source, obscured from direct visibility by an optically opaque circumstellar disc, is shown in the top panel, obtained with the HST WFPC2 at a waveband centered at 8140 \AA. The angular scales of each of the three frames are indicated on their right-hand sides. The orientation and width of the two CTIO spectrograph slits are indicated on both the bottom and middle panels, and are labeled CTIO 1 and CTIO 2, respectively.\label{fig2}}
\end{center}
\end{figure}

%\section{OBSERVATIONS}
We obtained low-resolution, long slit spectra of the RR with the R-C (Cassegrain) spectrograph at the Cerro Tololo Inter-American Observatory (CTIO) 1.5 m telescope. These observations were made on March 26, 2003, using a 2.5'' wide slit, 7.7' long. The grating \#09 with 300 l mm$^{-1}$, blazed at 4000 \AA, provided a 8.6 \AA\ resolution and a spectral coverage of 2600 \AA. Using a CuSO$_4$ filter to select the grating's first order, the setup covered a wavelength range of 3400-6000 \AA\, including the range from H$_\beta$ to the Balmer discontinuity. The Loral IK CCD detector yielded a spatial scale of 1.3'' pixel$^{-1}$ along the slit. We used a coronographic decker assembly to minimize scattered light from the star while probing as much of the inner nebula as possible. All observations were made using the full extent of the 7.7' long slit to get simultaneous sky observations. We obtained spectra at two nebular locations, 2.5'' and 5'' south of the central star HD 44179 with the slit in E-W direction. The nebular exposures were bracketed by exposures of the central star. Individual exposures were limited to 5-10 minutes on the nebula and 1 minute for the star, and 4-5 exposures were obtained for each orientation. Data Reductions were carried out with IRAF 2.12 EXPORT, and all spectra were flux calibrated via observation of standard stars.

\section{RESULTS}
\subsection{Line-depth Technique} 
We used the line-depth technique to initially detect and measure the fluorescence spectrum at each of the two nebular positions. This technique allows the identification and measurement of any continuous emission in a reflected spectrum, such as represented by the blue spectrum of the RR, relying upon the comparison of the depths of nebular and stellar spectral absorption lines. It works particularly well with the strong hydrogen Balmer lines at 6562, 4861, 4340, 4101, 3970, 3889, 3835, 3798, 3770 \AA ..... In the absence of any emission processes such as continuous fluorescence by PAH molecules, the relative line-depths in the reflected spectrum would be identical to the corresponding ones in the spectrum of the illuminating star, assuming that the wavelength-dependent scattering properties of the dust, e.g albedo and phase function asymmetry,  remain unchanged over the few-angstrom width of each individual line. Presence of any continuous emission manifests itself in that the lines in the reflected spectrum have smaller line-depths than those in the illuminating source spectrum. The illuminating source in this case is the star HD 44179, having an ideal spectral type A with strong Balmer lines. Also, at T$_{eff}\leq$ 9000 K the star is not likely to produce Balmer Line emission in the surrounding nebula. The decrease in the depth of a line can be directly related to the amount of underlying continuous emission (in this case, the fluorescence intensity). As the star is enshrouded by the disk and as the stellar spectrum is actually seen in reflected light, if there were any fluorescence in this zero-offset spectrum, the line-depth technique would reveal any additional fluorescence at the offset positions in the nebula. 

As an example we show in figure 3 the normalized spectra at the zero offset position and at an offset 2.5''S, 3.9''E of H$_\zeta$ at 3889 \AA. One can see that the line-depth at the nebular position is substantially diminished.

%place fig3 here
\begin{figure}
\plotone{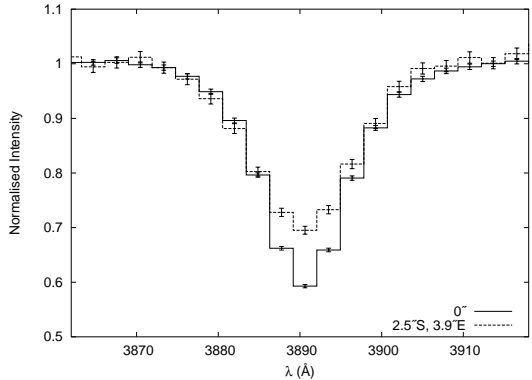}
\caption{Normalized spectra from two positions, zero offset and 2.5''S, 3.9''E at 3889 \AA, illustrating the filling in of the Balmer line due to fluorescence. Error bars are photon-statistics errors.}
\end{figure}

In the presence of any continuous emission the relative line depth in the nebula $R_N=(C_N - L_N)/C_N$ is smaller than the relative line depth $R_S=(C_S - L_S)/C_S$ of the same line in the illuminating star, where $C$ and $L$ refer to continuum and line center intensities, respectively, and where the subscripts $N$ and $S$ refer to the nebular and stellar spectrum, respectively. The fluorescence intensity relative to the scattered nebular continuum can therefore be expressed as $R_S/R_N - 1$ \citep[see][for a review]{wv04}.

Given the nature of our measuring technique, we can determine the fluorescence intensity only at the wavelengths of the resolvable Balmer lines as well as the Balmer discontinuity, i.e. at ten wavelength positions with our spectral resolution in our spectral range. By measuring the line-depths of each of the Balmer lines in the star and the nebula, we extracted the fluorescence intensities at those wavelengths. Table 1 shows the fluorescence intensities at the two offsets: 2.5''S, 3.9''E and 5''S, 6.5''E. Independent observations by the authors using the 90'' Bok telescope (Steward Observatory) and the B\&C spectrograph show corroborating evidence for this UV/blue fluorescence in the RR (U. Vijh, A. Witt, \& K. Gordon, in preparation). 

%place tab1 here
\begin{deluxetable}{cll}
\tabletypesize{\footnotesize}
\tablecolumns{3}
\tablewidth{0pc}
\tablecaption{Fluorescence Intensity at two positions in the RR nebula.}
\tablehead{
\colhead{Wavelength} & \multicolumn{2}{c}{Intensity (erg cm$^{-2}$ s$^{-1}$\AA$^{-1}$sr$^{-1}$)}\\
\cline{2-3}\\
\colhead{} & \colhead{2.5''S, 3.9''E} & \colhead{5''S, 6.5''E}
}
\startdata
4861.3  &4.059$\pm0.026\times10^{-6}$ &9.528$\pm0.145\times10^{-8}$\\
4340.5  &1.209$\pm0.008\times10^{-5}$ &1.571$\pm0.025\times10^{-6}$\\
4101.7  &2.914$\pm0.020\times10^{-5}$ &7.032$\pm0.123\times10^{-7}$\\
3970.1  &3.565$\pm0.023\times10^{-5}$ &3.441$\pm0.048\times10^{-6}$\\
3889.1  &3.817$\pm0.024\times10^{-5}$ &3.164$\pm0.045\times10^{-6}$\\
3835.4  &3.517$\pm0.022\times10^{-5}$ &1.881$\pm0.029\times10^{-6}$\\
3797.9  &4.438$\pm0.024\times10^{-5}$ &3.892$\pm0.047\times10^{-6}$\\
3770.1  &5.428$\pm0.023\times10^{-5}$ &5.483$\pm0.048\times10^{-6}$\\
3749.8  &6.417$\pm0.019\times10^{-5}$ &5.666$\pm0.034\times10^{-6}$\\
3570.0  &1.000$\pm0.019\times10^{-5}$ &1.180$\pm0.034\times10^{-6}$\\
\enddata
\tablecomments{Errors are combined photon-statistics errors.}
\end{deluxetable}

\subsection{Identification}
Hydrogenated amorphous carbon (HAC), silicon-nanoparticles (SNP), SiC grains, and small PAHs are all known to luminesce at blue wavelengths and have also been considered as likely candidates to explain the ERE \citep{wv04}. Figure 4a shows fluorescence spectra of HAC, SNP and SiC \citep{li00,huang02,liao02,belo00,pat00}. It reveals that the observed BL in the RR is unlike the luminescence spectrum of HACs, SNPs, or SiC. Given the exceptionally strong UIR-band emission in the RR, the most likely source of the newly discovered BL may then be fluorescence by PAH molecules. The nebular BL spectrum suggests a peak in the fluorescence intensity corresponding the first electronic transition $S_1$ - $S_0$ near 3750 \AA. Comparison with Figure 1 suggests PAH molecules in the mass range of 170 to 270 amu as the likely sources. Shown in Figure 4b are comparisons with laboratory fluorescence spectra \citep{chi01a,chi01b,reyle00} of gas-phase PAHs in that mass range with the BL spectrum of the RR at an offset of 2.5''S, 3.9''E. 

%place fig4 here
\begin{figure*}
\plottwo{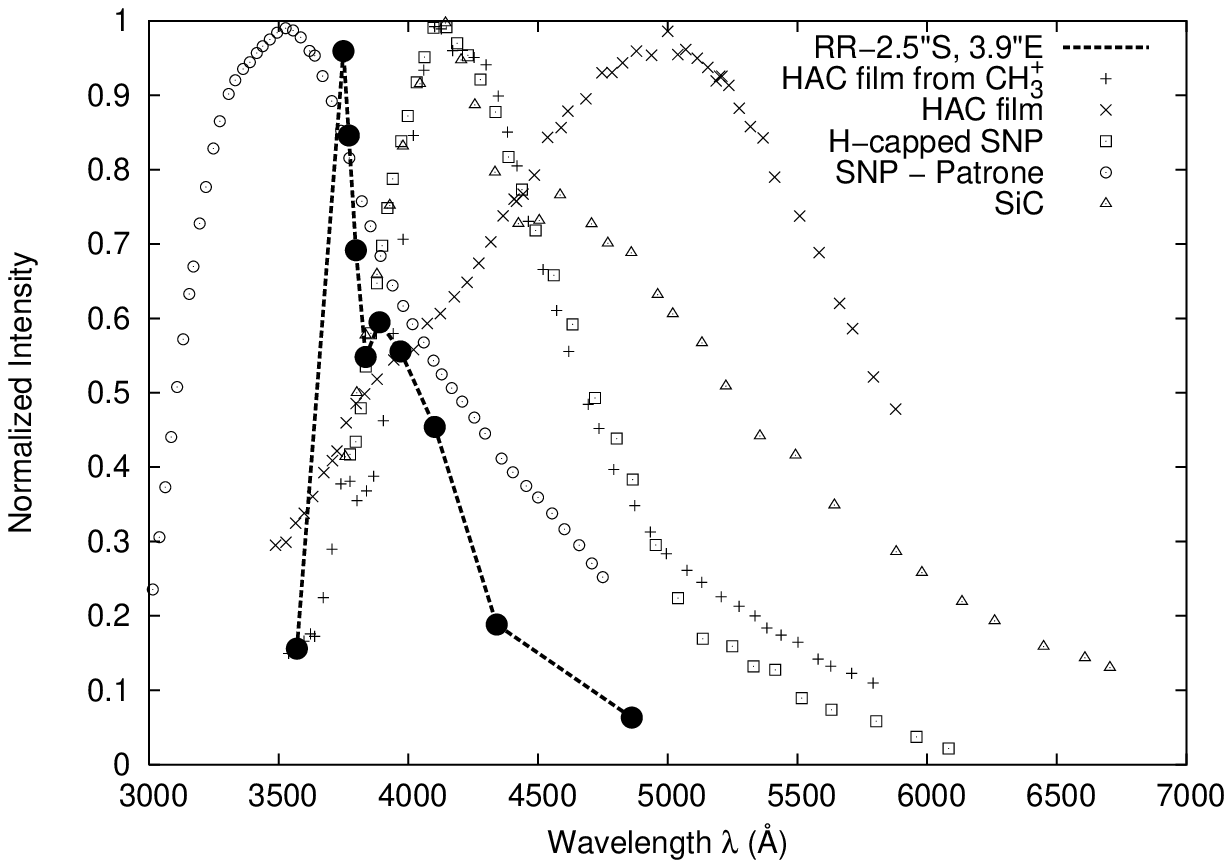}{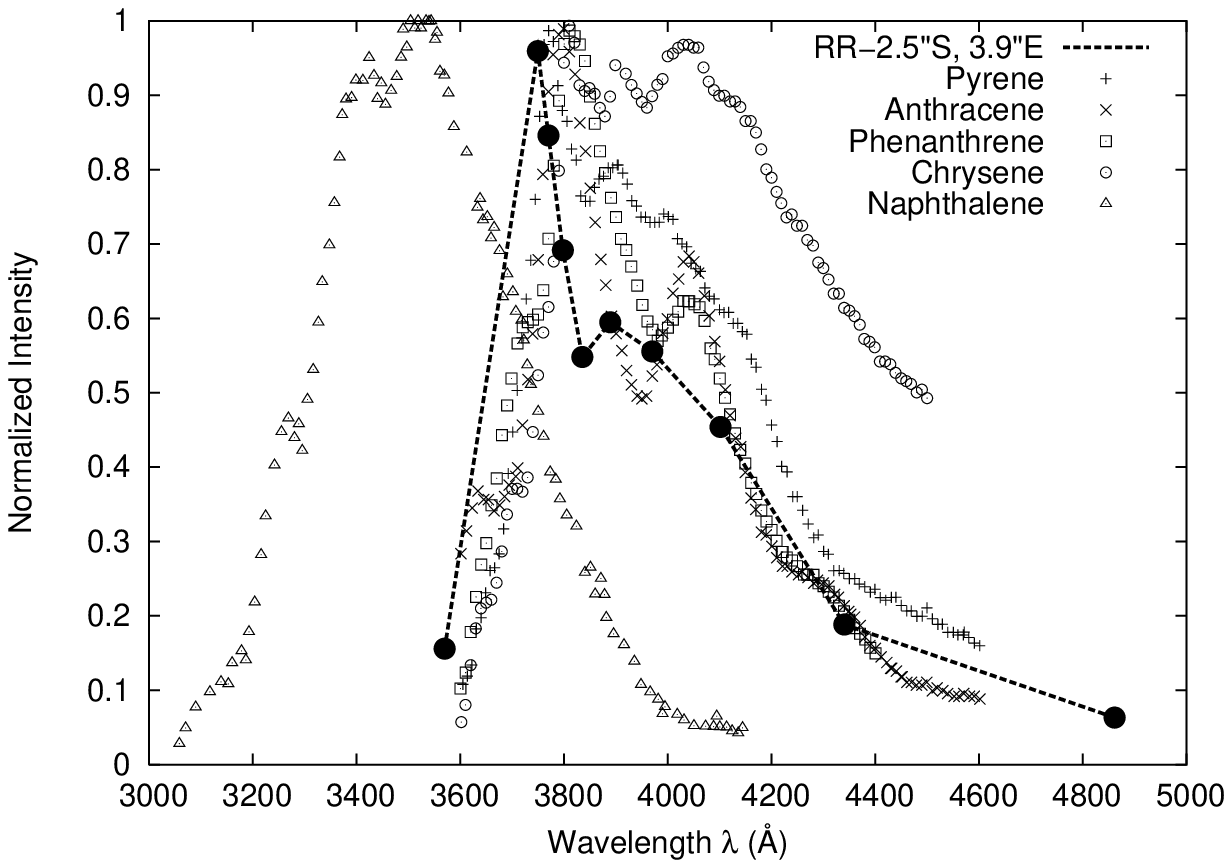}
\caption{The UV/blue fluorescence spectra of the RR nebula compared to PAH and solid-state fluorescents. (a), comparison with some solid-state candidates and (b), comparison with small neutral PAHs. The intensities are normalized for easy comparison of the spectra with existing laboratory fluorescence spectra. The dashed lines representing the RR fluorescence spectrum are drawn to guide the eye (measurements exist only at specific wavelengths indicated by solid circles).}
\end{figure*}

%The largest known interstellar molecule thus far is the 13-atom  molecule HC$_{11}$N, a chain molecule. 
Laboratory spectra obtained with PAHs in the gas-phase are more likely to be comparable to astrophysical spectra. These gas-phase spectra show considerable variation with temperature, becoming smoother, losing band-structure and the peak shifting to longer wavelengths with increasing temperatures \citep{chi01a,chi01b}. Though anthracene suggests a good match to the observed spectrum, we do not have sufficient spectral resolution to completely rule out other possible candidates, as the measurement technique works only at the discrete wavelengths of the Balmer lines.

\section{DISCUSSION}
Another important factor which should be considered when identifying possible carriers is fluorescence efficiency. Although many different fluorescing species may exist with similar abundances, the spectra of the most efficient ones will be dominant. Laboratory studies have shown that on comparison of emission rates per molecule, from PAH molecules placed at 1 AU from the Sun, anthracene and pyrene show 10-100 times more efficiency than naphthalene (C$_{10}$H$_8$) and phenanthrene (C$_{14}$H$_{10}$) \citep{brech94}. Additional support for small PAHs as sources of the observed BL is found in the spatial correlation between the BL and the PAH 3.3 $\mu$m emission in the RR and the presence of the distinct PAH ionization discontinuity in the spectrum of the central source, HD44179 (U. Vijh. A. Witt, \& K. Gordon, in preparation).

\acknowledgments
We are grateful to David F. Malin for providing a rare blue image of the RR nebula. UPV acknowledges a CTIO thesis student travel grant. This research was made possible through a generous allocation of observing time at CTIO and through grants from the US National Science Foundation and NASA. We would also like to thank Dr. Louis Allamandola for his valuable comments on the manuscript. We thank the anonymous referee for his/her comments which have significantly improved this paper.

%\clearpage

%\clearpage

%\clearpage 

%\clearpage

%\clearpage

%\clearpage

\end{document}